\documentclass[conference, compsoc]{IEEEtran}
\IEEEoverridecommandlockouts
\usepackage{cite}
\usepackage{amsmath,amssymb,amsfonts}
\usepackage{algorithmic}
\usepackage{graphicx}
\usepackage{textcomp}
\usepackage{xcolor}
\usepackage{multirow}
\usepackage{tabularx}
\usepackage{xcolor}
\usepackage{hyperref}

\newcommand{\codename}{COGENT}
\newcommand*{\CHERI}[1]{\textsc{CHERI}}
\newcommand*{\ZERO}[1]{\textsc{ZER\O}}
\newcommand*{\PUMP}[1]{\textsc{PUMP}}
\newcommand*{\RetTag}[1]{\textsc{RetTag}}
\newcommand*{\TimberV}[1]{\textsc{Timber-V}}

\newcommand{\ignore}[1]{}

\usepackage{listings}
\lstdefinelanguage[RISC-V]{Assembler}
{
  alsoletter={.}, %
  alsodigit={0x}, %
  morekeywords=[1]{ %
    lb, lh, lw, lbu, lhu,
    sb, sh, sw,
    sll, slli, srl, srli, sra, srai,
    add, addi, sub, lui, auipc,
    xor, xori, or, ori, and, andi,
    slt, slti, sltu, sltiu,
    beq, bne, blt, bge, bltu, bgeu,
    j, jr, jal, jalr, ret,
    scall, break, nop
  },
  morekeywords=[2]{ %
    .align, .ascii, .asciiz, .byte, .data, .double, .extern,
    .float, .globl, .half, .kdata, .ktext, .set, .space, .text, .word
  },
  morekeywords=[3]{ %
    zero, ra, sp, gp, tp, s0, fp,
    t0, t1, t2, t3, t4, t5, t6,
    s1, s2, s3, s4, s5, s6, s7, s8, s9, s10, s11,
    a0, a1, a2, a3, a4, a5, a6, a7,
    ft0, ft1, ft2, ft3, ft4, ft5, ft6, ft7,
    fs0, fs1, fs2, fs3, fs4, fs5, fs6, fs7, fs8, fs9, fs10, fs11,
    fa0, fa1, fa2, fa3, fa4, fa5, fa6, fa7
  },
  morecomment=[l]{;},   %
  morecomment=[l]{\#},  %
  morestring=[b]",      %
  morestring=[b]'       %
}
\def\BibTeX{{\rm B\kern-.05em{\sc i\kern-.025em b}\kern-.08em
    T\kern-.1667em\lower.7ex\hbox{E}\kern-.125emX}}
\begin{document}

\title{Generic Tagging for RISC-V Binaries}
\author{\IEEEauthorblockN{David Demicco, Matthew Cole, Gokturk Yuksek, Ravi Theja Gollapudi, Aravind Prakash,\\ Kanad Ghose and Zerksis Umrigar\\}
	\IEEEauthorblockA{\textit{Department of Computer Science} \\
		\textit{State University of New York at Binghamton}\\
		Binghamton, New York, USA \\
		\{ddemicc1,mcole8,gokturk,rgollap1,aprakash,ghose,umrigar\}@binghamton.edu}
}

\maketitle

\begin{abstract}
With the widespread popularity of RISC-V \--- an open-source ISA \--- custom hardware security solutions targeting specific defense needs are gaining popularity. 
These solutions often require specialized compilers that can insert metadata (called tags) into the generated binaries, and/or extend the RISC-V ISA with new instructions.
Developing such compilers can be a tedious and time-consuming process.
In this paper, we present \codename{}, a generic instruction tag generator for RISC-V architecture. 
\codename{} is capable of associating a tag of configurable and varying widths (1 to 20 bits) to each instruction.
It is also capable of emitting labels that are central to the implementation of control-flow integrity (CFI) solutions.
\codename{} encodes all tags and labels as nop instructions thereby providing full backward compatibility. 

We evaluate \codename{} on a subset of programs from the SPEC CPU2017 benchmark suite and report the binary size increase to be 29.3\% and 18.27\% for the lowest and highest tag coverage levels respectively. 
Additionally, we executed tagged programs on Commercial Off The Shelf (COTS) RISC-V unmodified hardware and found the execution time overhead (w.r.t. backward compatibility) to be 13.4\% and 5.72\% for the lowest and highest coverage levels respectively. 
Finally, using a case study, we present possible use case scenarios where \codename{} can be applied. 
\end{abstract}

\section{Introduction}\label{sec:introduction}
Over the years, the advent of Internet of Things (IoT) along with an increasing number of mobile, desktop and server systems has led to a steady increase in cybersecurity attacks.
Most of the attacks--even those targeted at the hardware (e.g. Spectre\cite{Kocher2018spectre}, and Meltdown\cite{lipp2018meltdown}) are initiated through software.

The defense community has focused efforts at different layers of deployments.
For instance, the software community has explored software-only defenses such as StackGuard\cite{cowan1998gaurd}, and CFI via program instrumentation (e.g., ~\cite{zhang2013control,niu2015per,abadi2009control}).
Meanwhile, the hardware community has pursued defenses that leverage hardware features in order to detect and flag exploits (e.g., Intel CET\cite{shanbhogue2019cet}, W\textasciicircum{}X) or prevent information leakage (e.g., Intel SGX\cite{intelsgx}, ARM TrustZone\cite{armtrustzone}).

While software-only security measures are highly desirable and can be deployed with little effort, they typically suffer higher performance penalties compared to hardware solutions. Furthermore, the security mechanisms themselves may be a part of the attack surface, which brings about transparency issues.
On the other hand, hardware-only security mechanisms must bridge the semantic gap problem and recover high-level security-relevant program semantics in order to enforce rich and expressive policy abstractions.
Further, the tolerance for errors is low and ``fixing" design errors is very expensive (e.g. Intel MPX\cite{Intel2013,Ramakesavan2016} had problems that resulted in eventual deprecation).

With the gaining popularity of RISC-V, an open-source ISA, hardware - software cohesive defenses have gained center stage and are desirable.
Typically, defenders take advantage of (a) fine-grained control over the RISC-V hardware including (but not limited to) changes to ISA in combination with (b) curated compilers capable of enriching binaries with security metadata/stubs, deploying highly effective and fast performing defenses.
For example, recent defenses such as Cheri~\cite{watson2015cheri}, Zero~\cite{ziad2021zero}, RetTag~\cite{wang2022rettag}, Framer~\cite{Nam2019}, etc., all rely on a dedicated compiler that can generate instrumented binaries for their modified hardware.
The modified binaries contain metadata called ``tags'' that provides the hardware with specific information regarding the code or data unit in the program.
Typically, there is a one-to-one mapping between a tag and corresponding code/data unit.
These defenses strike a balance between security and performance requirements and are often tailor-made to deployment needs.
They rely on performant compilers and/or hardware design and engineering, yet compiler development remains an expensive process often requiring several hundred developer-hours for development and testing.
More importantly, availability of a suitable compiler would attract more hardware defenses that are currently impeded by the lack of a compiler.  
Solutions that shorten the development effort and allow quick incorporation of metadata into binaries are missing and highly sought after.

In this paper, we address the above need and present \codename{}: COmpiler for GENeric Tagging, a generic LLVM-based compiler that incorporates instruction tags (or code metadata) into RISC-V binaries.
In essence, {\em \codename{} removes the burden of compiler development from RISC-V hardware defenses that rely on embedding instruction metadata into binaries.}
Motivated by the needs of Control-Flow Integrity (CFI) defenses that embed CFI labels into code, \codename{} interleaves tag metadata into the instruction stream.  
\codename{} has several key features that make it highly favorable for hardware-software cohesive defenses.
First, it is highly {\em flexible}.
It can accommodate various tag widths (1 to 20 bits per instruction).
This flexibility is essential in order to achieve rich expressiveness in tags.
Second, it is {\em functionally correct}.
That is, it ensures that the inserted tags do not interfere with program logic.
Functional correctness is non-trivial due to several challenges (see Section~\ref{subsec:overview:challenges}).
This is important since a tag system is not usable without such guarantees.
Third, \codename{} preserves backward compatibility.
That is, tagged binaries will seamlessly run on generic RISC-V processors that are tag-unaware.
We accomplish backward compatibility by encoding tag information into NOP instructions that are simply ignored by unsupported hardware.
Finally, our changes are segregated into LLVM IR-level and target-specific changes.
Other architectures (e.g., ARM) can be supported by reusing the IR-specific changes and rewriting ARM LLVM backend.

We make the following contributions:
\begin{itemize}
    \item {We implement a flexible and defense-scheme agnostic implementation codenamed \codename{}, a LLVM-based compiler for RISC-V target. \codename{} provides configurable instruction-level tagging support to several different tag widths and coverage. We also make changes to the LLD linker to support \codename{}.}
    \item{We demonstrate the efficacy of \codename{} via a CFI implementation.}
    \item{We evaluate the example CFI implementation for correctness and backwards compatibility. We demonstrate moderate to low performance overhead and binary bloat.}
    \item Through case studies, we demonstrate potential real-world applicability of \codename{}.
\end{itemize}

\vspace{.04in}
\noindent
{\bf Summary of Results:}
We evaluated the correctness of \codename{}{} by compiling and running a subset of the SPEC CPU2017 benchmark suite on Commercial Off The Shelf (COTS) hardware. 
Our subset has small function sizes, heavy memory overhead, and high loop counts, all of which stress instruction caches. 
For our example CFI implementation, we found that \codename{} increased the size of the {\tt .text} section of these programs by an average of 29.32\% for the lowest tag coverage, and 18.27\% for the highest. 
We found that backward-compatibility performance (i.e., overhead imposed by tagged binaries running on tag-unaware hardware) was impacted by an average of 13.40\% for the low coverage and 5.72\% for the highest. 
We also present several possible use cases for the system, including specific recent CVEs that could be addressed.

The rest of the paper is organized as follows: ~\autoref{sec:background} provides the technical background necessary to understand the remainder of the paper.
In \autoref{sec:overview} and \autoref{sec:technical_details}, we present the overview and technical details of \codename{}.
We evaluate \codename{} in \autoref{sec:eval} and provide case studies to highlight benefits of using \codename{}.
Finally, we present related work in \autoref{sec:related_work} and conclude in \autoref{sec:conclusions}.

\section{Background}\label{sec:background}
\subsection{Security via Tagging}
    Tagging is a technique that very broadly associates some amount of metadata (the tag), with a memory location and registers. Specially designed architectures then consume that tag and enforce a security policy. Tags can be associated with a) memory objects (data tagging) allowing policies to track and protect data in memory, and/or on b) code (instruction tagging), allowing policies to change or extend the behavior of the tagged instructions. Of particular importance to this paper is instruction tagging.
    
    \vspace{.04in}
    \noindent
    {\bf Instruction Tagging:} allows a tag-aware architecture to execute additional code or checks on any instruction associated with a particular tag.
    Instruction tags are useful in implementing Control-Flow Integrity (CFI) policies.
    Identical tags are used at control flow sources (call/jump sites) and control flow targets in order to signify legal control flow.
    
    Tags can be placed on load/stores to protect sensitive data, forbidding access to a memory region if the tag is incorrect. These kinds of tags often require corresponding metadata placed with the data they are meant to access. 
    The \textit{tag policies and enforcement are implementation specific and form the essence of a tag-based defense}.
    The specific instruction tags are a function of the intended security policies, and therefore vary greatly between designs \cite{dhawan2015architectural}.
    Differences in tag requirements can even result in different designs which use the same hardware for similar purposes \cite{nick2018protecting}.
    This implementation-specific design requirement is inherent to the nature of tagging. Tags are simply a way of providing more information to the program at run time, but the information needed is dependent on the design of the system itself.
    \par Broadly speaking, there is a trade-off made in the design of tag systems between how much information it provides to the hardware, and how much overhead the tags introduce into the system. This overhead can include power consumption, hardware complexity, performance or any other costs paid by the design. Depending on how much information one's defense needs, tags can be arbitrarily large, or as small as a single bit. Designers often use techniques such as encoding the metadata information into the unused upper bits of a pointer, but this is often insufficient on its own \cite{wang2022rettag}, requiring additional tagging or ISA(instruction set architecture) changes. The choices of where to store metadata, how much metadata to store, and how to access it are defining characteristics of tagging systems.
    
\subsection{LLVM Compiler Infrastructure}
    The LLVM compiler infrastructure \cite{chris2004llvm} is a fully featured compiler designed with modularity in mind. It is broken down into components, which operate on different parts of the compilation process. Important to \codename{} is the idea of the language-specific front end, the target architecture-specific back end, and the language- and target-agnostic intermediate representation (IR). The LLVM design requires the language-specific front end to convert source code into LLVM IR, the optimizer schedules an analysis and transformation pass pipeline on that IR, and its output is IR for consumption by the back end. The back end might also perform additional transformations on the code via its own set of passes and built-in functionality. The final product of the back end is then output to an object file, which is fed into a linker to create the final binary.
    Modifying the LLVM compiler can thus take two main forms. The simplest scenario is to create and modify transformation and analysis passes within the existing structure of the pipeline. These transformation passes can be scheduled to run before or after any specific pass or combination of passes, allowing one to place new transformations and analysis at any point in the pipeline. The more complex scenario is to modify the internal components of LLVM, changing or extending the functionality of pre-existing code in the LLVM infrastructure.

\subsubsection{Relocations and Symbols}
    As part of the creation of an object file, the back end will emit target specific information called a \textit{relocation} into object files. These relocations are designed to inform the linker of how to handle information that cannot be finalized before the linking phase of compilation. Examples of this for RISC-V  include encoding addresses, which are often emitted as a pair of relocations (named HI and LOW), covering two instructions (i.e. \texttt{auipc, jalr}). These relocations tell the linker to place the final address of the target symbol across two instructions when it is required. Symbols themselves can be kept as offsets from an unknown starting address, allowing the linker to know the final address during the linking process, even though the absolute position cannot be determined by the compiler back end.

\section{Overview}\label{sec:overview}
\subsection{Goals}
Tagging schemes are complex and varied, and different schemes require different configurations. There are some that require more space (i.e., metadata) to be accessible for any given instruction, while some might require only a single bit for some defenses \cite{ziad2021zero}. Additionally tagging schemes must be able to provide the metadata to the hardware. A generic instruction tagging scheme must be able to meet this requirement under as many different configurations as possible. To that end we set out to create \codename{}.

The goal of \codename{} is to enable embedding configurable metadata tags into instruction stream, and not implementation details for different defense schemes such as forward or backward edge CFI, heap or stack, stack protection, or any other particular defense scheme. We did not set out to create a specific defense with this paper, or provide analysis that can be used to create one. Instead we focused on three major tasks. 

\begin{itemize}

\item {\emph{Flexibility:} \codename{} must be able to generate binaries for RISC-V architecture that require varying tag widths, with minimal changes. 
Additionally, we aim to modularize our implementation in order to minimize the developer effort in modifying \codename{} to support a new architecture. 
\codename{} must also be able conducive to supporting different types of defense schemes including those that require large amounts of metadata.}
\item {\emph{Functional Correctness:} Insertion of tags and metadata must not interfere with program logic. 
That is, all necessary tags must be appropriately inserted, and inserted tags must not alter the control flow or corrupt data.} 
\item {\emph{Backwards Compatibility:} The tool must allow created code to be backwards compatible. 
That is, tagged binaries must be able to seamlessly execute on tag-unaware hardware. While backwards compatibility is a desirable feature for many systems, we recognize that it may not be necessary in every case. If turning it off could provide benefits to a scheme \codename{} should allow that as well. }
\end{itemize}

\subsection{Challenges}\label{subsec:overview:challenges}
\subsubsection{Metadata Lookup costs} A tag or metadata lookup action in the hardware could be expensive and inefficient. Therefore it is best to reduce the number of abstractions required to retrieve any tag for a given instruction. Likewise having separate (but mapped) memory adds complexity to the process of bringing the tags into the hardware. %
    
\subsubsection{Tag Semantic Gap} 
During the compilation process, programs pass through different stages of compilation. 
While the tags are eventually assigned on the raw instruction stream, the actual assignment of tags may occur at a higher level of abstraction (i.e., source code or intermediate representation). 

Depending on the stage of compilation where tags are assigned, percolating them down to the instructions is important and requires bridging of the semantic gap between the higher and lower levels of program representation. 
Further, during compilation the compiler will often expand IR instructions/source statements into multiple assembly instructions. This manifests as a \emph{Many-to-One} relationship where many tags map to a single IR instruction/source statement. A very common example in RISC-V is the lowering of the LLVM-IR pseudo instruction PSEUDOCALL from MachineInstr level into the MCinst level. 
During this process the single PSEUDOCALL is expanded into an \texttt{auipc,jalr} pair. 
Likewise there are \emph{One-to-Many} relationships where multiple instructions at a higher-level of abstraction are lowered into a single assembly instruction (e.g. bitcast instructions). 
These one-to-many and many-to-one relationships can occur during any point of the lowering process, and where possible, \codename{} must provide ways to handle them correctly. Unfortunately there are many cases without a generic solution i.e. in the lowering from IR level instructions to machine Instructions. 

\subsubsection{Compiler-inserted Code} During compilation the compiler may introduce instructions that did not exist at the previous (i.e., higher) level of representation. This is typically done to meet back end specifications or optimizations. Examples include in-lining and function prologue/epilogue insertion. \codename{} must provide a way to tag compiler-inserted code, and must ensure that it does not disrupt the tag schemes. 

\subsubsection{Pre-calculated Offsets, Symbols, and Relocations} Features such as C++’s try/catch block implementation often pre-calculate the distance between two symbols instead of leaving it as a relocation wherever possible. \codename{} must be sensitive to these pre-calculated values. More generally, wherever possible the compiler calculates the offset to a relocation or a symbol from the beginning of a compilation unit. 
\codename{} must preserve the relative locations in the code section though any changes it makes. 

\subsubsection{ISA Changes} Prior efforts have altered the ISA to enrich code with metadata.
Any ISA changes would by their very nature break backwards compatibility. 
\codename{} must be able to encode tags without any modifications to the ISA.

\subsection{Our Approach}
When examining both the challenges and the goals, we converged on the following design choices for \codename{}. These solutions address both the challenges and goals. 

\subsubsection{Flexibility} To address flexibility and the many back-end specific challenges we encountered, we designed \codename{} to be as modular as possible, with front end and back end components in order to allow for adaptability. Unfortunately many (but not all) of the back end changes are target specific in the best case, and design specific in the worst. To solve this we clearly separate the \codename{} process into five major pieces.

\begin{itemize}
\item{\emph{IR Level Tagging API}. This component allows the creation of IR level passes that can take advantage of the front and back end agnostic nature of the IR to perform analysis and select tags for IR level instructions. These IR level tagging passes can be shared between any source and any target, and so are the most portable method of tag selection. This component consists of an analysis pass that provides propagation of assigned tags into any back-end passes where possible.}

\item{\emph{Backend Level Tagging API}. This component allows the creation of LLVM MachineInstr and MCinst level passes that can tag code after it has been lowered into a target specific back end. This allows tagging of code inserted after IR level optimizations or back end insertions.}

\item{\emph{Backend and Defense Specific Changes}. These changes are both back end and defense specific. Representing choices that must be made in how the target back-end and the scheme must interact. For the RISC-V prototype we developed (see \autoref{sec:example}), One example would be the insertion of a metadata label between an \emph{auipc jalr} pair to include a CFI label. A second would be the specific behavior of tags in the many to-one and one-to-many lowering relationships. These changes cannot be handled by back end passes alone and must be hard coded into the back end itself on a case-by-case basis.}

\item{\emph{Defense Scheme Agnostic Back End Changes}. These are changes to the back end that must be done to allow for the insertion of the tag instructions as late as possible into the code stream. This includes turning off the pre-calculated offsets, ensuring correct tag placement, and adjusting symbol and relocation offsets before object file emissions. These do not interact with the logic of tag selection, only with the emission of the tags themselves into an object file.}

\item{\emph{Configuration File}. This file allows a user to specify simply what emission options they would like for the final tag instructions. Details of the provided options are presented in \autoref{sec:select}, but the configuration file allows developers to switch between tag widths for their own instruction tagging schemes.}
\end{itemize}

To address the flexibility requirement of providing additional metadata where needed, we introduce metadata labels as a tool for a developer to use. These labels can be repurposed for any use, and are encoded as nops by the compiler in order to maintain backwards compatibility. In the RISC-V back end we chose \texttt{lui} instruction which provides 20 bits of metadata per inserted label. 

\subsubsection{Functional Correctness}.
Addressing correct tag placement becomes challenging when inserting new tag  instructions at specified alignments. Any changes made to the code after a tag has been inserted will displace the tags and must be adjusted in order to maintain consistent tag placement. To this end, all functions are forced to be aligned.
Additionally, we assume that relaxation/compressed instruction generation is turned off (see \autoref{sec:compressed} for a discussion on compressed instructions). Additionally optimizations such as pre-calculating offsets based on the MCInst’s as described in challenges \emph{must} be turned off. This adds additional one time overhead to the linking process, but no extra run time cost beyond the insertion of the tags themselves. Optimizations introduced by IR passes can be left in place however, by insuring tag insertion is completed at the last point possible in the pipeline. Finally tag instructions can be placed in locations that disrupt relocation pairs, meaning the linker must be made aware of this possibility. 

\subsubsection{Backward Compatibility}
To allow for backwards compatibility, at least one of the possible tag instruction configurations must be encoded as a nop instruction. For the RISC-V prototype, the instruction {\tt lui x0, <imm>} was selected, as it is functionally a {\tt nop} and gives the maximum number of bytes to encode tag values. Other target architectures would have to select different nop-equivalent instructions with differing amounts of bits to ensure backwards compatibility.

\begin{table*}[ht]
    \centering
    \begin{tabular}{|l|c|c|c|c|c|c|c|}
        \hline
        {\bf Instruction}   & \multicolumn{1}{l|}{{\bf Bits available}} & \multicolumn{1}{l|}{{\bf C1 bits}} & \multicolumn{1}{l|}{{\bf C3 bits}} & \multicolumn{1}{l|}{{\bf C7 bits}} & \multicolumn{1}{l|}{{\bf C15 bits}} & \multicolumn{1}{l|}{{\bf C31 bits}} &  \multicolumn{1}{l|}{{\bf Backwards compatible?}} \\ \hline
        {\tt addi}          & 12        & 12        &           4   & 1     & NA & NA   & Yes                                       \\ \hline
        {\tt lui}           & 20        & 20        &           6   & 2     & 1   & NA  & Yes                                     \\ \hline
        Custom opcode & 25        & 15        &           8   & 3     & 1   & NA   & No                                 \\ \hline
    \end{tabular}
     \caption{Tag instructions used by \codename{}.}\label{fig:taginstructions} C1, C3, C7, indicate the number of instructions "Covered" by one tag instruction. Coverage 1 Coverage 3, Coverage 7 etc. Higher coverage's mean less tags in the instruction stream. Bits available indicates the total amount of metadata space for that instruction, and the Bits columns shows how much metadata there can be per instruction.  
\end{table*}

\section{Details}\label{sec:technical_details}
With the solutions outlined in \autoref{sec:overview} we created a prototype version of \codename{}. This prototype comes in two parts, The underlying changes to LLVM itself, and an example implementation that provides metadata labels on computed jumps and computed callsites.

\subsection{LLVM Changes}
We make all of the changes described in this section to LLVM version 10.0.0.

\subsubsection{Tag Instruction Selection}\label{sec:select}
    The first and the most important step in \codename{} implementation is to provide a way for the developer to select how they wish to represent the tags in the instruction stream. This representation is referred to as the tag instruction. The selection of this tag instruction determines
\begin{itemize}
    \item How many bits a user will have available to store tags
    \item How many instructions any single tag can cover
    \item Backwards compatibility
\end{itemize}
    \autoref{fig:taginstructions} shows the different choices we provide via a configuration file.
    We detail their bit layout, how many bits they provide, and the number of bits available for each instruction based on the coverage desired. If backwards compatibility is not required, users can select a custom opcode for the tag instruction.
    The tag enumeration itself is defined in a separate header file that must be included to the relevant passes.

\begin{figure*}[ht]
\centering
    \includegraphics[scale=.8]{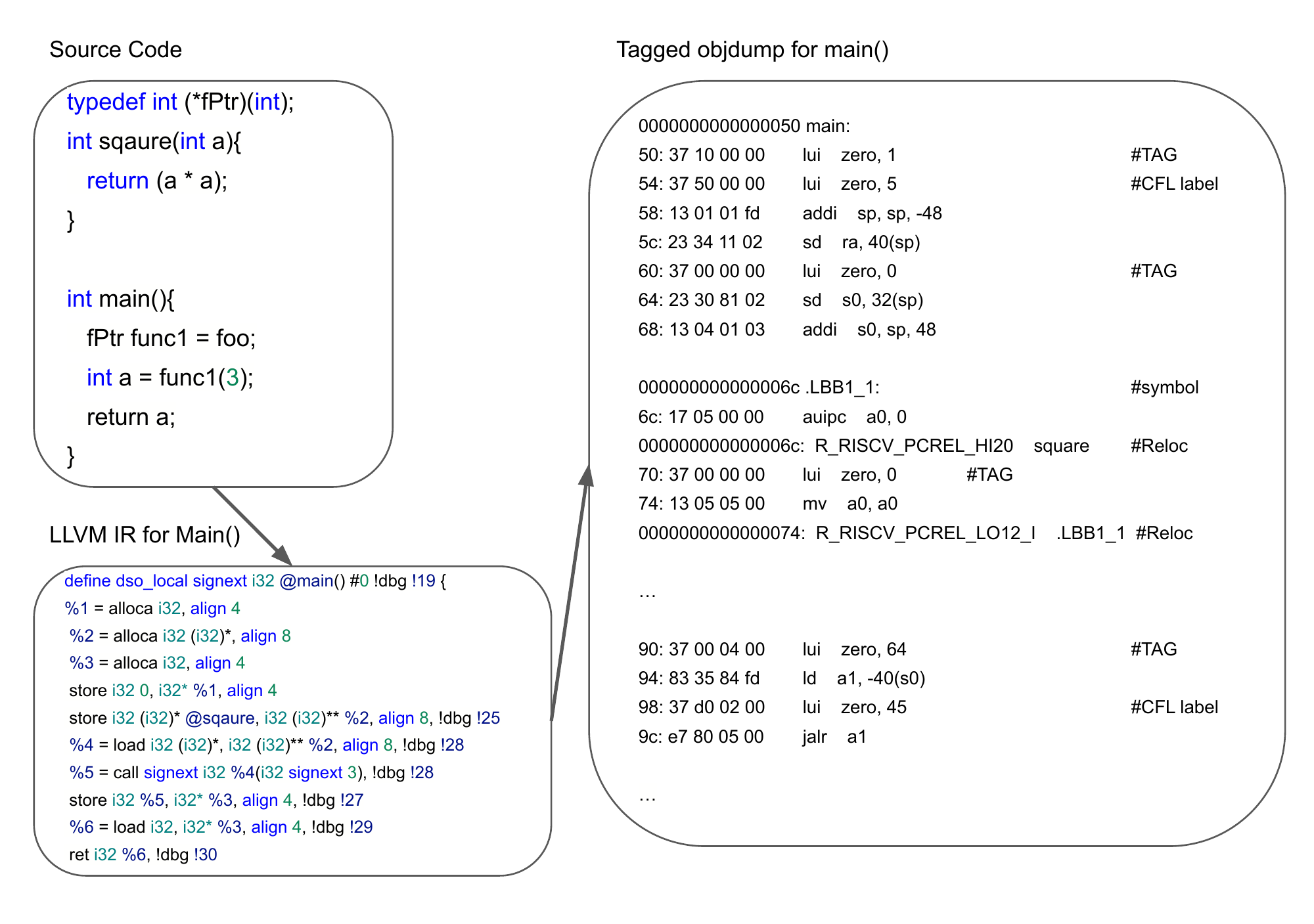}
    \caption{Demonstration of \codename{} using the simple CFI described in \autoref{sec:example}, contains compiler inserted code (0x58-0x68), many to one relationships (0x94-0x9C), relocation and symbol adjustments (0x6c-0x74), a tag inserted between a relocation (0x70).  Additionally it includes data that must be persisted from the IR (the function labels) to the backend inserted labels at (0x50) and (0x98).}
    \label{fig:cfiexample}
\end{figure*}

\subsubsection{Tagging API}\label{sec:tagAPI}
    Once the tag instruction has been selected, the next \codename{} process is to assign tags to LLVM’s IR level instructions or machineInstr. In the backend, this process is done via a simple API we expose to the user, consisting of standard {\tt .set(<tag>)} and {\tt .get(<tag>)} extensions on top of LLVM’s base machineInst classes. These are employed by LLVM backend passes that developers insert into the compiler pipeline. In general tag instrumentation passes should be inserted as late as possible in the pipeline to prevent optimizations from interfering with tag lowering. We also provide label emission, via an emission of a single {\tt lui} nop at any point in the compilation process. This {\tt lui} must be tagged appropriately to indicate to the hardware to process this label as a particular type of metadata.
    
    For tagging in the IR level, we provide a pass that IR level passes can push a mapping of {IRInstruction : tag} into, and that backend passes can register with to access that tag mapping to assign those tags to machineInstr. The lowering of these IR tags is complicated by the needs of the target specific backend, as described in the next section.  

\subsubsection{Lowering Behavior}
For IR instruction tagging, we provide a default behavior in a modified RISCV backend.
However depending on the specific deployment needs, developers must investigate the lowering process and make changes where appropriate to their planned tagging behavior.
By default one-to-many transformations apply the tag on the IR instruction on every machine instruction created by that instruction during the lowering process. For many-to-one, the default behavior is the \emph{first} tag. Any instructions not directly created from another instruction (an optimization or specification specific to the back end) will be assigned the default tag value from the tag configuration file.
This process may be insufficient for specific defense needs, but there is no one-size-fits-all approach to the lowering problem and necessitates developer effort.
    For machine instruction inserted tags, we modify the lowering process from machineInstr to MCInst allowing the tag field to be copied over. In cases where this is a one to one lowering, we simply copy the field from the machineInstr to the MCInst. For one to many relationships, such as PSEUDOCALL, we assign multiple tags to the machineInstr which are assigned to each of the MCInst’s during this expansion process. Specific tag values for a given expansion will depend on the tagging scheme itself. Similar to IR level lowering there is no one size fits all solution to expansion. Fortunately at the machineInstr level there are only a few of these cases to handle in the RISCV back-end.

\subsubsection{Emitting the Tags}
The actual tag emission into the instruction stream occurs in the {\tt finalize} function during code emission in the LLVM backend.
    At this point the byte code for instructions has been placed into containers called fragments, and their layout relative to each other is known. However they have not been emitted into the final binary. We modify LLVM such that as instructions are encoded into their fragment, their ordered list of tags are also added into a new “tag” array.
    With this tag array in place, and the knowledge of the fragments offset (provided by llvm) we create a new data array for the fragment, that has the tag instructions inserted into the instructions.
    We do this for every fragment that contains instructions. \emph {In this manner we can be sure that any MCinst that has received a tag, will have that tag emitted into the final binary}
    Any non-zero fragment offset must be recalculated to account for tags that will have been inserted into previous fragments. However because this is a simple one tag instruction per-number of encountered instructions, it is always possible to calculate this correctly with just the original offset. (I.E. If the offset is 12 bytes, and the tag coverage is 3, there would be a single tag instruction inserted before that offset, and the new offset would then be 16)  
    Like the fragment offsets, during tag emission symbols and relocation must be adjusted to account for the tag insertions. As they contain offsets that are byte distances from the beginning of a fragments, these values must be adjusted by the same one per n calculation as the fragment offsets. For both symbols and fragments we create a new array, with the updated values, and then overwrite the unmodified arrays using a ConstCast.
    There are some corner cases in symbol resolution that must be addressed. Specifically the compiler will attempt to calculate the distance between two symbols before the finalize layout step if it is able to determine the distance early (i.e., if they are in the same fragment), this is a compiler optimization to increase symbol resolution speed in the linker. To deal with this, we disable this optimization and allow the linker to resolve the difference after tag insertion.
    Additionally, the calculation of if a jump should be PC-relative or indirect is done before the final emission based on the maximum distance allowed in a single PC-relative jump instruction. This calculation must be adjusted to assume an additional one tag instruction per number of instructions insertion. An example of the final result of these changes, and the final emission of tags is presented in \autoref{fig:cfiexample}

\subsubsection{Linker Changes}
    To support both the insertion of tags and possibly metadata labels, the linker must be made aware that there can be instructions inserted between a HI and LO relocation pair. Normally HI-LO relocation's must appear as a pair, directly next to each other. We modify the linker to allow for up to three instructions inserted between the pair, and adjust the final value calculations to account for the added distance. Three additional instructions are selected as a default to allow HI, tag, METADATA\_LABEL, METADATA\_LABEL, LO configurations, however this number can be adjusted if more metadata is desired.

\subsection{Example CFI Implementation}\label{sec:example}
    We provide an example CFI implementation that demonstrates the common design challenges and changes of a simple \codename{} implementation. CFI is a very well studied problem \cite{abadi2009control}\cite{Dewey:2012:CFI}\cite{yuan2020cracks} and is a common use case for tagging implementations \cite{dhawan2015architectural}\cite{ziad2021zero}\cite{wang2022rettag}. This example CFI implementation places metadata labels at the beginning of exported or address taken functions, and before the function call at computed call sites. It consists of the following components.  
\begin{itemize}
\item{A tag configuration with 2 tags, using a lui-nop to provide backwards compatibility. The tags are simply N (A normal instruction) or CFL (control flow label) this allows any of the coverage's supported by \emph{lui} \autoref{fig:taginstructions} to be selected.}
\item{An IR level pass that assigns CFI labels to functions and callsites based on their function signature. Function-signature-based CFI policy is consistent with the CFI approach in LLVM.
A callsite and a destination must have matching CFL labels.}
\item{A Machine instruction level pass that inserts and tags the CFI labels with CFL at the beginning of every non-internal function.}
\item{A back-end and scheme specific change to insert the CFL label between the AUIPC-JALR pair that RISCV generates for call sites.}
\end{itemize}
An example of this CFI implementation on a from source to object file is provided in \autoref{fig:cfiexample}. This example also includes many of the common challenges faced by \codename{}.

CFI is such a well studied problem, and our example solution is based on previous work \cite{chris2004llvm}. As such we do not claim this implementation as a novel contribution. 

\ignore{
Hardware support for this scheme would consist of checking that the instruction after a CFL label is a jump, and the first non-tag instruction encountered after that jump is a matching CFL label. Of particular note is the back-end and scheme specific change. Not every scheme would need this label inserted before the call, and so this is not a generic component, and serves as a prime example of the scheme and target specific nature of many-to-one and one-to-many relationships.}

\subsection{Further Expansions}
\subsubsection{Tag Expansion}
Although we do not claim it as a contribution, there is no fundamental reason that placing two tag instructions next to each other would not work. Such a configuration could allow for higher coverage at the cost of higher space overhead. This may be useful when space is a nonissue, or for inserting complex metadata for testing purposes.
\subsubsection{Compressed Instruction Support}\label{sec:compressed}
Likewise there is no fundamental barrier to support compressed instructions.
The instruction opcode in RISCV indicates if an instruction is compressed or non-compressed.
With that knowledge and the current PC, it is possible to infer what bits in the tag instruction contain the corresponding tag.
For compressed instruction support, care must be taken when designing the tags and the tag policy, as the tag bits must be small enough such that a single tag instruction can cover every instruction (compressed or not), even if they are all tagged. This can be done by using tags small enough to cover all possible instructions (i.e., 3 bit tags for an {\tt lui} with an uncompressed coverage of 3) or a two-tiered tag system, where uncompressed instructions receive a larger share of the tag space. The compiler must additionally be modified to insert alignment NOPs in the case where a non-compressed instruction would cross a tag instruction, pushing it under the coverage of the next tag.

\subsubsection{Dynamic Linking}
Currently \codename{} only supports statically linked binaries. This is to ensure that the tags for any externally linked libraries are always located in the correct place for interpretation. We plan to support dynamic linking in future versions of \codename{}.

\subsubsection{Handwritten Assembly Support} Handwritten Assembly sections are loaded into the compiler backend for parsing before being emitted to the final object file. This allows \codename{} to insert the tag instructions in hand written assembly segments. However its ability to correctly determine tags on these sections is limited. The correct tagging of these sections would be left to the implementation of a specific tagging scheme, in much the same way as backend and scheme specific tagging would be and may only be possible with manual tagging.

\subsubsection{Additional Configurations}
Other tag layouts beyond the single one we chose are desirable in a generic tagging system. Possibilities include
\begin{itemize}
    \item Placing all the tags for a function's instructions at the beginning of that function. This would consist of a series of tag instructions sufficient to cover every instruction in the function. On non tag aware systems this would simply be the equivalent of a series of NOPs that run before a function starts.  
    \item Placing the tags for a basic block at the beginning of that basic block. Again, this would consist of a series of tag instructions large enough to cover every instruction in that basic block.
    \item Placing the tag a separate .text.tag section in the ELF executable file. It would be up to the hardware to determine how to use this section, but it can be desirable, and non tag aware hardware can simply ignore the section completely.
\end{itemize}

While \codename{} currently does not support these additional configurations, we have plans to extend support to them in the near future.  

\begin{figure*}[ht!]
    \centering
    \includegraphics[scale=.75]{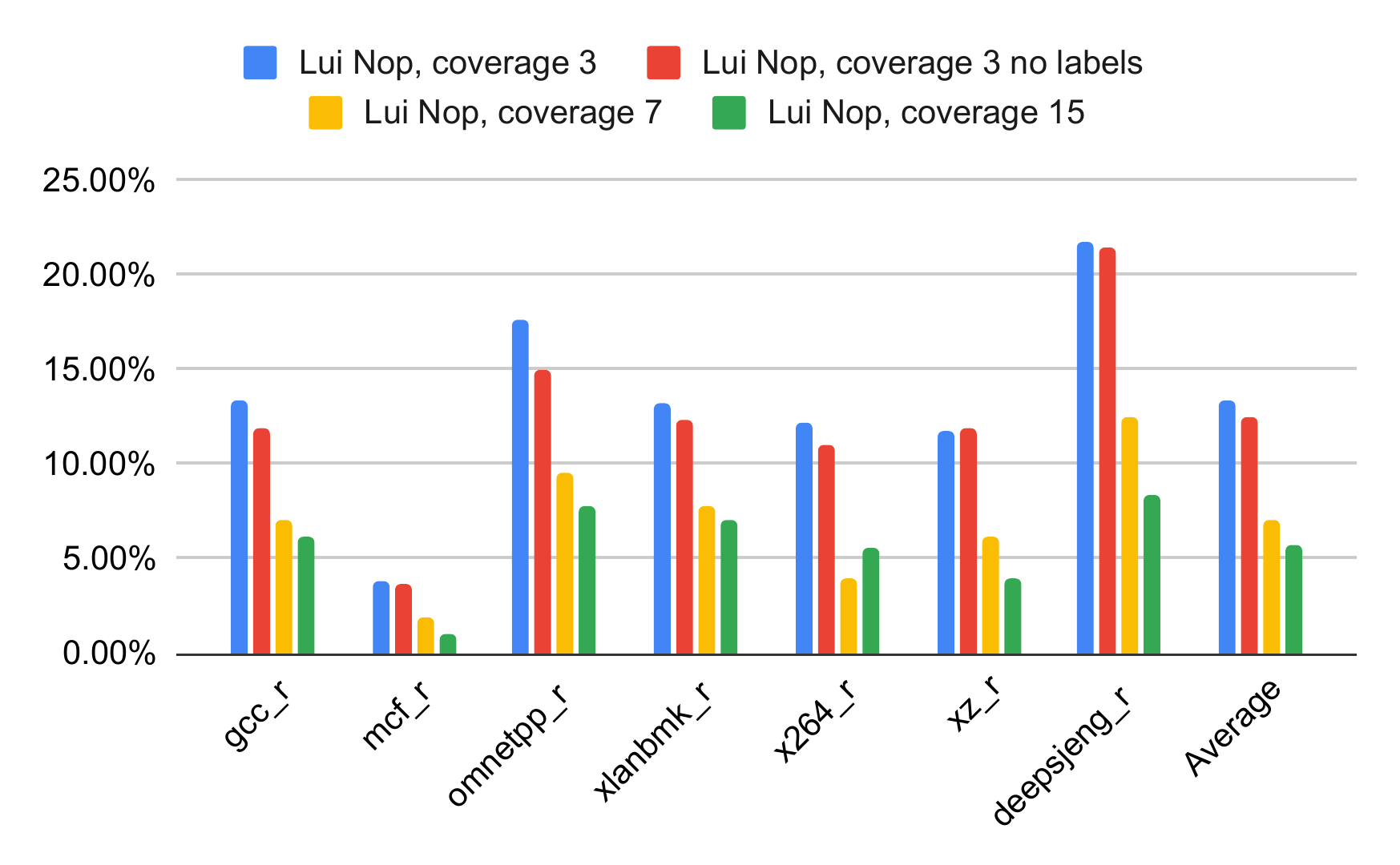}
    \caption{Backward compatibility results for Lui NOP, coverage 3, 7, and 15, with and without labels in percentage overhead}
    \label{fig:overhead}
\end{figure*}
 \section{Evaluation} \label{sec:eval}

\codename{} is a generic system with many different possible use cases. As such we evaluate it Four primary ways. Overhead on COTS hardware (i.e. the overhead of backwards compatibility.), and binary size overhead, are presented to demonstrate the cost of cogent in its most common configurations and uses. The last two, are a comparison of related works, and case studies demonstrating how the system can easily be adapted to different use-cases. 

The off the shell performance tests were done on a SiFive U540 SoC, which contains four 64-bit RISC-V cores and 8GB of memory, running Gentoo Linux with the kernel version 5.2.9.

For the backwards compatibility evaluation, we use the {\tt lui} nop described in \autoref{sec:select}. This is chosen to allow the maximum amount of space per tag and still allow the binaries to run on unmodified hardware. Coverage 3 provides 6 tag bits per instructions covered, 7 provides 2, and 15 provides one. While the labels were inserted as described in the example CFI configuration (see \autoref{sec:example}), evaluating them requires special hardware capable of interpreting the tags. This is explored in a separate hardware paper \cite{gollapudi2023star}. 

We chose a subset of the SPEC CPU2017 rate benchmarks for our tests. This subset has a wide range of characteristics, including heavy memory operations in mcf\_r, and rapid arithmetic operations in deepsjeng\_r \cite{ranjan2019spec}. Testing against these characteristics reveals what kind of performance impacts can be expected on unmodified hardware. SPEC CPU2017 provides three different types of workloads. Test, training, and reference. The test workload is a simple test of the executable, the training is designed to give feedback for optimizations, and the reference workload is the fully timed data set. For these tests we used the full reference workload.

The size overhead includes the overhead of label insertion on the coverage 3 configuration. This is done comparing the size of a binary compiled with and without labels for each benchmark program. A simple count of the number of inserted labels would be insufficient when including tags, as the position and number of labels can add to the number of tags inserted into the program.

\subsubsection{Backwards Compatible Results}
    Overall the average backward-compatibility performance overhead of inserting a tagged {\tt lui} nop every three instructions and labels was 13.40\%. Every 7 instructions and labels was 6.98\%, and every 15 instructions and labels was 5.72\%. This is not a linear reduction due to two major factors. The inclusion of the labels, whose overhead is the same over every coverage, and varied properties of benchmarks themselves. Additionally, for small loops and functions in binaries the number of nop’s encountered in the code stream can be greater than simply 1 per 3 normal instructions.
    
\begin{lstlisting}[
	basicstyle=\small\ttfamily,
	caption=Sample RISC-V assembly loop,
	label=asm-ex]
 loop_start:
    <TAG>
    addi
    addi
    addi
    <TAG>
    j    loop_start
\end{lstlisting}  

Consider the simplified toy example in \autoref{asm-ex}. In this set of instructions, we encounter 2 tags every 4 normal instructions every time we iterate over the loop. Examples can be found for every coverage size, using larger loops or functions.

\subsubsection{Binary Properties}
\label{binaryProperties}
    The properties of the benchmarks have a major impact on the results of our evaluation. For programs that are memory operation intensive, even in backwards compatibility mode, the overhead of additional nops added to the instruction stream is quickly overwhelmed by the cost of memory operations. mcf\_r has this property, and has an overhead of 3.79\% for coverage 3, and 1.09\% for coverage 15 in backwards compatibility mode. Conversely, small looping in-register operations can cause the nops to have an oversized impact on program performance. Deepsjeng fits this profile, and the resulting overhead is 21.79\% for C3, 12.45\% for C7, and 8.39\% for C15. Programs that do not exhibit such extremes in behavior tend to have overhead results close to the average results. Full results can be found in \autoref{fig:overhead}.

\subsubsection{Size Overhead}
    Overall we expect the text section to increase in proportion to the number of tag instructions we insert. However complicating this number is the function alignment requirements. In the text section of the binary, the space between functions is filled in with zeros that extend from the end of one function to the beginning of the next. This effect is more pronounced in C++ programs with many small functions such as omnetpp\_r and xlancbmk\_r. Labels have a minimal impact on binary size, averaging 0.78\%. Overall the text section size increase is on average 29.32\% for coverage three with labels, 20.49\% for coverage seven, and 18.27\% for coverage 15. The full text section size results are available in \autoref{tab:size}.

\begin{table*}[ht]
\centering
\begin{tabular}{|l|r|r|r|r|r|}
\hline
                 & \multicolumn{1}{l|}{Baseline} & \multicolumn{1}{l|}{luic3} & \multicolumn{1}{l|}{luic3 no label} & \multicolumn{1}{l|}{luic7} & \multicolumn{1}{l|}{luic15} \\ \hline
gcc\_r O0       & 14,872,236                    & 21,237,048                 & 21,082,660                          & 18,689,424                 & 17,984,840                  \\ \hline
mcf\_r O0       & 99,900                        & 143,022                    & 141,850                             & 125,626                    & 121,722                     \\ \hline
omnetpp\_r O0      & 4,242,910                     & 6,126,344                  & 6,007,300                           & 5,565,696                  & 5,536,696                   \\ \hline
xlanbmk\_r O0      & 8,057,876                     & 11,430,065                 & 11,190,441                          & 10,452,873                 & 10,447,205                  \\ \hline
x264\_r O0         & 858,148                       & 1,189,080                  & 1,181,316                           & 1,042,292                  & 998,772                     \\ \hline
xz\_r O0           & 317,973                       & 440,498                    & 436,670                             & 391,798                    & 379,986                     \\ \hline
deepsjeng\_r O0 & 217,512                       & 307,464                    & 305,860                             & 269,684                    & 259,092                     \\ \hline
\end{tabular}
\caption{Text section size in bytes for baseline and 4 different configurations. Luic3 is coverage 3 using the lui nop. luic7 is coverage 7 using the lui nop. luic15 is coverage 15 using the lui nop.}
\label{tab:size}
\end{table*}

\begin{table*}[ht]
\centering
\begin{tabular}{|l|p{1.5cm}|p{1.5cm}|p{1.5cm}|l|p{1.5cm}|p{3cm}|}
\hline
 & \textbf{Flexibility} & \textbf{Variable Tag Size} & \textbf{Arbitrary Metadata} & \textbf{ISA Changes} & \textbf{Uses Pointer Bits} & \textbf{Memory Protection} \\ \hline
\PUMP{}\cite{dhawan2015architectural} & High & No* & Yes** & Yes & No & Tags only readable by handlers \\ \hline
\CHERI{} \cite{watson2015cheri} & Low & No & No & Yes & Fat pointers & Capability registers \\ \hline
\RetTag{} \cite{wang2022rettag} & Low & No & No & Yes & Yes & hardware protected secure key \\ \hline
\ZERO{} \cite{ziad2021zero} & Low & No & No & Yes & Yes & Must protect tagged memory \\ \hline
\TimberV{} \cite{weiser2019timberv} & Low & No & No & Yes & No & Must protect tagged memory \\ \hline
\codename{} & High & Yes & Yes & No & No & W\textasciicircum{}X protection \\ \hline
\end{tabular}
\caption{State of the art comparison. *\PUMP{}'s tags are \texttt{sizeof(int *)} per memory word, some applications will not use the full space. **\PUMP{} allows arbitrary metadata by allowing a pointer to be encoded into the tag}
\label{tab:comp}
\end{table*}

\subsubsection{Feature Comparison Against State of the Art}
\codename{} was created to enable flexibility in its deployment. To demonstrate this against the current state of the art, we compared \codename{} against recent tag-based architectures using several criteria.
\begin{itemize}
    \item {How easy is it to employ the tool in a different manner then one of the creators designs? A high rating means the tool was designed for that purpose, while a low indicates it was created for one specific task.}
    \item{Does this tool have a Variable tag size?}
    \item{Can this tool provide an arbitrary amount of metadata to a user above and beyond the size of the tag, if that user desires it.}
    \item{Does this tool change the underlying ISA, removing backwards compatibility.}
    \item{Does this tool use the unused pointer bits?}
    \item{What kind of additional memory protection does this tool employ to protect its tags?}
\end{itemize}
The results of this investigation are available in \autoref{tab:comp}. 
In general we find that unless a tool was designed for reuse (i.e. \PUMP{}), it would be difficult to re-target the hardware for other tasks. 
State of the art solutions use ISA changes to provide the compiler with the wrappers and tag aware instructions they need to manipulate their metadata. 
These ISA changes break backwards compatibility. The common use of unused pointer bits, or custom pointers, render solutions incompatible with each other. 

\begin{table*}[ht]
\centering
\begin{tabular}{|l|l|l|l}
\cline{1-3}
\textbf{CVE} & \textbf{Description} & \textbf{\codename solution} &  \\ \cline{1-3}
CVE-2021-3330 & Several points of failure lead to a large buffer overflow & Type Aware Arithmetic &  \\ \cline{1-3}
CVE-2022-24310 & An integer underflow leads to a large buffer overflow & Type Aware Arithmetic &  \\ \cline{1-3}
CVE-2022-23613 & A missing lower bound check leads to an integer underflow & Type Aware Arithmetic & \\ \cline{1-3}
\end{tabular}
 \caption{Sample CVE's and tag-based solutions}
 \label{tab:CWE_CVE}
\end{table*}

\subsection{Case Studies and Use Cases}
\codename{} is a generic tool designed to allow for the easy implementation of other tagging and labeling schemes. We present a case study, offering a detailed look at systems \codename{} could be used to achieve. We selected CVE-2021-3330 to explore with detailed solutions presenting and describing Tag Aware Arithmetic \autoref{sec:2021-3330}. \autoref{tab:CWE_CVE} shows a selection of recent CVEs to which our findings could be applied, including the one we selected for our case study.
Additionally, we present an alternate usage of \codename{} for fuzzing and testing, where we use \codename{} to keep track of coverage and branching behavior of a program at runtime.

\begin{table}[ht]
\centering
\begin{tabular}{|l|l|l|}
\hline
\multicolumn{1}{|l|}{\textbf{Tag}} & \multicolumn{1}{l|}{\textbf{Enum}} & \multicolumn{1}{l|}{\textbf{Meaning}} \\ \hline
N & 0 & \begin{tabular}[c]{@{}l@{}}Instruction should not be checked for\\ Integer over/underflow
\end{tabular} \\ \hline UN\_ARTH & 1 & \begin{tabular}[c]{@{}l@{}}This instruction should be checked for\\ integer overflow/underflow\end{tabular} \\ \hline

\end{tabular}
 \caption{Overflow check tagging scheme}
 \label{tab:overflow_check}
\end{table}

\subsection{Case study: CVE-2021-3330}\label{sec:2021-3330}
\subsubsection{CVE Description}
    CVE-2021-3330 is a vulnerability that affects the Zephyr RTOS, and comes from a series of errors that includes an improperly constructed cyclical list, where a singularly linked list is expected. Iterating over this now cyclical list causes an unsigned value to be repeatedly subtracted from another unsigned value, leading to an integer underflow. This value is then fed as the length argument into \lstinline[language=C]|memmove()|, leading to a massive memory move that can overwrite kernel structures. This CVE consists of several points of failure. Improper list construction, integer underflow, and the overwrite itself. In this case study, we focus on a potential solution for the integer underflow, which would prevent the memory overwrite by triggering an exception when the underflow occurs. 
\subsubsection{Type Aware Arithmetic}
    To do this we will introduce type aware arithmetic, requiring a tag with a width of a single bit, giving two possible states for every instruction as shown in \autoref{tab:overflow_check}. These two tags are N and UN\_ARTH. The N tag is given to any instruction in the stream that will not be checked for integer overflows. This can be because the instruction does not perform any arithmetic on an unsigned integer, or because the programmer or the compiler has chosen to disable the check on that instruction. While the UN\_ARTH would be automatically placed on any instruction that performs arithmetic operations on unsigned types.
    To insert this tag, an llvm compiler pass would be created that examines the left hand side of any arithmetic expression, determining if the type of the result is unsigned. If it is, the compiler will add the tag to any and all arithmetic instructions that result from the IR level instruction, which the \codename{} system would then propagate down into the final binary. There are two cases where the compiler would not place the UN\_ARTH tag on an instruction that creates an unsigned value. The first is when the compiler can determine that values being subtracted are hard coded immediate values, in this case we would have to assume the programmer is attempting to exploit the defined behavior of unsigned (IE unsigned MAX = 0 - 1). The second would be a provided annotation that allows the programmer to \emph{opt out} of the checks if they had a reason to. This is required as in the C standard integer over/underflow are defined behavior. Forcing the user to opt out of the behavior is done in an attempt to minimize developer errors when using this defense. 
\subsubsection{Required Hardware Support}
    The required hardware support for this tag scheme is very simple, if an instruction is tagged with UN\_ARTH, an exception should be thrown if the results from that operation over or underflow. Hardware would require this additional tag support as without metadata, it would be unable to differentiate between an operations on signed and unsigned integers. This can either be a hard stop to the program, or be call an exception handler if the policy calls for it. In this case, for CVE-2021-3330, we would halt the program execution, as the corruption that results from this bug can overwrite kernel data structures. 

\subsubsection{Code and Tag Layout} 
For CVE 2021-3330, the original integer overflow occurs in the following two lines of code.

\begin{lstlisting}[language=C++, 
    frame=single,
    caption={CVE-2021-3330 underflow C++ source}]
memmove(frag->data, frag->data +
    hdr_len, frag->len - hdr_len);
frag->len -= hdr_len;
\end{lstlisting}

These lines of code operate only on unsigned integers, and do not have the opt-out annotation. After compilation, this code produces the following relevant lines of assembly, loading the required values into registers, and then subtracting them. 

\begin{lstlisting}[
    linerange={1-10}, 
    breaklines=true,
    frame=single,
    caption={CVE-2021-3330 underflow ASM}
]
lhu  a2, -34(s0)  #Load hdr_len into a2
...
lhu  a0, 8(a0)    #load frag->len into a9
sub  a2, a0, a2   #subtract frag->len - hdr_len
...
call memmove      #calls memmove
lh   a1, -34(s0)  #loads hdr_len into a1
ld   a2, -32(s0)  #loads frag into at2
lh   a3, 8(a2)    #loads frag->len into A3
sub  a1, a3, a1   #subtracts frag->len - hdr_len
\end{lstlisting}

There are two places in this code that will need to be tagged, \emph{sub a2, a0, a2} and \emph{sub a1, a3, a1}. This information can be determined by the compiler, and so they will receive the UN\_ARTH tags. Resulting int he following final code layout. Note the tag instructions themselves are not included in this listing, as there exact position would depend on compiler optimizations levels, and overall layout of the code, however they would be inserted into the instruction stream at the appropriate alignments. 

\begin{lstlisting}[
    linerange={13-22}, 
    breaklines=true,
    frame=single,
    tabsize=2,
    caption={CVE-2021-3330 underflow ASM}
]
lhu  a2, -34(s0)	[N]
...
lhu  a0, 8(a0)	  [N]	
sub  a2, a0, a2	  [UN_ARTH]
...
call memmove	    [N]
lh   a1, -34(s0)	[N]
ld   a2, -32(s0)	[N]
lh   a3, 8(a2)	  [N]
sub  a1, a3, a1	  [UN_ARTH]
\end{lstlisting}

With all this in place, when this tagged code is run and the bug is triggered, operation would proceed normally until the subtraction that causes the arithmetic underflow. 
Once that happens, the hardware would trigger an exception based on the UN\_ARTH tag, causing an extra check as part of the instruction processing phase.
A final strength of this solution is that because of the simple nature of the two tags employed, it would be a trivial exercise for \codename{} (\emph{although not necessarily the hardware}) to layer this tag with another single bit tag that would not be placed on arithmetic instructions. 
An example could be the single bit return address protection provided by \cite{dhawan2015architectural} and \cite{ziad2021zero}. These two protections could be layered without the need for extra tag sizes.  

\subsection{Capturing Code Coverage via Tagging}
\subsubsection{Description}
    Inline tagging could be used during testing and fuzzing to allow for quick tracking of program coverage, counts of reached instructions, and paths taken through the CFG. In this use case, run time security is not considered as part of the design. Instead \codename{} inserts metadata informing testing hardware what information it needs to record, and where it needs to record it. Hardware changes are limited to what is required to facilitate these tasks.

\begin{table*}[t!]
\centering
\begin{tabular}{|l|l|l|}
\hline
\textbf{Tag} & \textbf{Enum} & \textbf{Meaning} \\ \hline
CL & 0 & Increments the label value of this instruction when encountered \\ \hline
CI & 1 & Tells the hardware to record a counter for this PC in memory. \\ \hline
BCF & 2 & Tells the hardware to record what the next PC is for the current PC in memory. \\ \hline
N & 3 & Normal behavior \\ \hline
\end{tabular}
\caption{Coverage tags summary}
\label{tab:coverage}
\end{table*}

\subsubsection{Tag and Label Design} This use case requires the coverage 7, 2 bit LUI labeling scheme. 
Giving 4 possible tags, each indicating a different behavior to the testing hardware.
Additionally we introduce counting labels, with the initialized value of 0 as the primary fast coverage tracking method.
An in depth description of each tag and the labels are as follows, with a summary provided in \autoref{tab:coverage}. 

\begin{itemize}
\item{CL (Counting label) - Indicates that this instruction is a Counting label. When this tag, and the associated label are encountered in the instruction stream. First the value contained in them is incremented by one, Then \emph{written back} into the label. This requires that the system be allowed to write values back into the code section. These will be placed at the beginning of every basic block.}

\item{CI (Counting Instruction) - Indicates that this instruction should be counted. When the instruction stream reaches this tag, it will check the current program counter against a mapping of \emph{PC : Int} to see if this instruction has been encountered before. If it has, it will increment the associated Int by one. If not, it will create a new entry in the map, with a value of one. This counting instruction is explicitly slower than the counting label, due to the need for the mapping lookup, but could be used for specific purposes.}

\item {BCF (Branching Control Flow) - Indicates a branching point in the instruction stream that needs to be tracked. This tells the hardware to check if the current PC is in the BCF mappings. If it is not, it creates a new entry in a \{PC : \{Next\_PC : Count\}\} mapping. After it has either found, or created, the entry for the current PC, the Hardware then checks if the \emph{next} program counter is in the \{Next\_PC: Count\} mapping, adding it if it does not find it, or increment the counter if it does. In this way the BCF tag builds a map of each branching instruction, the targets it has reached, and how many times it has reached those targets.}

\item{N (Normal) - This tells the hardware that no special hardware actions are needed to enforce the overall security policies within this instruction.} 
\end{itemize}
\subsubsection {Compiler Support}
    For coverage tracking, the compiler would be configured to place counting labels and their associated tags at the beginning of every basic block at the machine instruction level. Additionally, a pass would be added to enable two levels of branching control flow tracking depending on the desires of the tester. BCF labels could be placed on every control flow instruction or, just on computed control flow instructions. 

\subsubsection {Hardware Support}
    This coverage system requires two major changes to the supporting hardware beyond the ability to read tags. First is the ability to write the incremented counters back into the counting labels. Second is the ability to record the PC of instructions tagged with CI or BCF into mappings. To do this we would reserve a known memory for the CI and BCF mappings, and ensure that the size of each entry is known to the hardware. This means for BCF mappings there would be a configurable maximum number of targets in order to keep lookup times reasonable. 

\subsubsection {Example Code} 
The RISC-V assembly code below is an example of how the coverage tagging would look like on a simple basic block, followed by a branching jump.

\begin{lstlisting}[
    linerange={1-19}, 
    breaklines=true,
    frame=single,
    tabsize=2,
    caption={Coverage Example ASM}
]
0x0		TAG	
0x4		lui		x0, 0		  			[CL]
0x8		lw		a1, -24(s0)			[N]
0xC		addi  a1, a1, 1				[N]
0x10	sw    a1, -24(s0)			[N]
0x14	lw    a1, -24(s0)			[N]
0x18	bne   a1, a0, .LBB0_2 [BCF]
.LBB0_1:
0x1C	lui		x0, 0		  			[CL]
0x20	TAG
0x24	lw    a0, -24(s0)			[N]
0x28	addi  a0, a0, -1			[N]
0x2C	sw    a0, -24(s0)			[N]
.LBB0_2:
0x30	lui		x0, 0		  			[CL]
0x34	addi  a0, zero, 5			[N]
0x38	sw    a0, -24(s0)			[N]
\end{lstlisting}

Assuming the above  was entered three times, jumping once and falling through twice, the counting label at 0x4 would be \lstinline|lui x0, 3| and at 0x1C it would be \lstinline|lui x0, 2|. 
LBB0\_2 is always reached at 0x34, making its value \lstinline|lui x0, 3|. 
The branching table would contain \texttt{0x18 : {0x1C: 2}, {0x30: 1}}, showing the exact behavior of the branch over three runs. 

\section{Related Work}\label{sec:related_work}
{\it \PUMP{}} \cite{dhawan2015architectural} provides a flexible system to associate metadata with memory locations, including on instructions. Its large fixed-width tag \emph{sizeof(*)} provides unlimited metadata size per memory location via allowing the tag to simply be a pointer to additional memory. However schemes that actually use this amount of space are not common, and even then, the common case for any given memory location is a small amount of metadata. To help alleviate the large overhead of having so much unused space in there tags, they propose and implement tag compression schemes to speed up retrieval. Compared to \PUMP{}, \codename{}'s instruction tagging allows simpler protection schemes to use smaller tag sizes when needed, while at the same time allowing that same arbitrarily sized metadata though the use of any number of inline labels. Importantly though, \PUMP{} has an actual hardware design backing it, while \codename{} does not have specialized hardware, so the only comparison can be between the location and storage size of metadata, not on performance, hardware complexity, or overall overhead.

\textit{\CHERI{}} \cite{watson2015cheri} is a well known capability based architectural enhancement for RISC instruction sets. Work on it is current and on-going \cite{WesleyFilardo2020} \cite{Davis2019} \cite{Xia2019}. CHERI provides its security though architectural capabilities, and tagged memory. Additionally it is a full stack solution, and includes compiler, OS, and hardware support. Cheri itself requires that compiler to support on all levels. Language specific front end, IR level passes, and backend emissions. 

{\it \RetTag{}} \cite{wang2022rettag} uses compiler modifications to insert new instructions in the function prologue and epilogue to generate a new instruction (thus modifying the ISA and preventing backward compatibility) called \emph{pac}. This instruction generates a unique PAC using the a unique function ID, the virtual address of the functions RA the SP, and a 128-bit RAA key and stores it in the 16 unused bits of a function pointer, and it is used to protect the return address from corruption. 

\textit{\ZERO{}} \cite{IbnZiad2021a}, provides a set of new instructions via a modification of the ISA and a method of encoding metadata that keeps the overhead to eight bytes per 4kb page. This work discards backwards compatibility by including new ISA instructions for memory and metadata management. Additionally it stores ten bits of information in the unused bits of pointers to enforce its own version of CFI using unique-per-function signature labels. It modifies the compiler to insert its new instructions at the appropriate places.

\TimberV{}~\cite{Weiser2019} allows the creation of up to four protected enclaves on a RISC-V system. It likewise uses a new set of instructions in order to tell the hardware when it can check and manipulate tags. Like \ZERO{} this removes its ability to provide backwards compatible code. While Timber-V does not \emph{require} compiler support, they specify that compiler support would allow for better integration. 

Several works including \textsc{Framer} \cite{Nam2019}, \RetTag{}~\cite{wang2022rettag} and \ZERO{}~\cite{ziad2021zero} use the upper unused bits of a pointer as place to store there metadata. Doing this gives these works strong backwards compatibility and low overhead. However by their nature it is difficult to layer any works that use the unused pointer bits. Additional space is limited, leading framer to use a novel encoding scheme to provide access to the pointers metadata, relying as much as possible on relative offsets from known points (i.e. \textit{frames}). Such works could be combined with \codename{}{} as we do not manipulate the pointers themselves, leaving the unused parts of the pointer available for any other scheme, or to be used for a CFI label.  

\textit{FlexFit} \cite{Delshadtehrani2021} is a generalized solution presented as a method to allow users to configure a set of instruction filters for use at run time. They implement this on a RISC-V FPGA board to demonstrate the practicality, feasibility, and usability of there generic filtering scheme. They store the protection information in four unused bits of the RISC-V PTE, giving up to 16 protected domains at a page granularity. Using the unused bits in the PTE, has been explored by other works as well \cite{delshadtehrani2021sealpk} \cite{shrammel2020donky}, which is why FlexFit limits its protection domains to 16, using only 4 of the available bits. While FlexFit does not use a compiler, the authors suggest a specialized compiler and loader for future work.

\section{Conclusion}\label{sec:conclusions}
This paper presents \codename{}, a modified LLVM compiler capable of generating RISCV binaries with configurable width instruction tags. 
These tags can be used to design security solutions that convey rich semantic information to a tag-aware architecture. 
\codename{} tackles several challenges including one-to-many and many-to-one instruction-level tag associations. 
We evaluate \codename{} on a subset of SPEC 2017 benchmark programs for performance and binary bloat. 
Further, we present case studies that highlight potential use cases of \codename{}. 

\bibliographystyle{IEEEtran}
\bibliography{tagging.bib}

\end{document}